\documentclass[a4paper,twoside]{jpconf-mod}
\usepackage{graphicx}
\usepackage[hyperindex,breaklinks]{hyperref}
\usepackage{url,amsmath,amssymb,cite,multirow} 
\def\tev{\textrm{TeV}}
\def\gev{\textrm{GeV}}
\def\pt{\ensuremath{p_{\rm T}}}
\def\met{\ensuremath{E_{\rm T}^{\rm miss}}}

\def\ifb{\ensuremath{\textrm{fb}^{-1}}}
\def\to{\ensuremath{\rightarrow}}
\def\X{\ensuremath{\tilde\chi_1^0}}

\def\t#1{\tilde{ #1}}

\def\mn{\ensuremath{\mu\nu}SSM}
\def\XX{\ensuremath{\tilde\chi_4^0}}
\DeclareGraphicsExtensions{.pdf}
\begin{document}

\markright{IFIC/15-07}

\title{Exploring neutrino physics at LHC via $R$-parity violating SUSY}

\author{Vasiliki A Mitsou}

\address{Instituto de F\'isica Corpuscular (IFIC), CSIC -- Universitat de Val\`encia, \\ 
Parc Cient\'ific de la U.V., C/ Catedr\'atico Jos\'e Beltr\'an 2, 
E-46980 Paterna (Valencia), Spain}

\ead{vasiliki.mitsou@ific.uv.es}

\begin{abstract}
$R$-parity violating supersymmetric models (RPV SUSY) are becoming increasingly more appealing than its $R$-parity conserving counterpart in view of the hitherto non-observation of SUSY signals at the LHC. In this paper, RPV scenarios where neutrino masses are naturally generated are discussed, namely RPV through bilinear terms (bRPV) and the ``$\mu$ from $\nu$'' supersymmetric standard model (\mn). The latter is characterised by a rich Higgs sector that easily accommodates a 125-\gev\ Higgs boson. The phenomenology of such models at the LHC is reviewed, giving emphasis on final states with displaced objects, and relevant results obtained by LHC experiments are presented. The implications for dark matter for these theoretical proposals is also addressed.
\end{abstract}

\markright{V~A~Mitsou}
%
\section{Introduction}\label{sc:intro}

Supersymmetry (SUSY)~\cite{susy} is an extension of the Standard Model (SM) that assigns to each SM field a superpartner field with a spin differing by half a unit. SUSY provides elegant solutions to several open issues of the SM, such as the hierarchy problem, the nature of dark matter~\cite{dm-review}, and the grand unification. SUSY is one of the most relevant scenarios of new physics probed at the CERN Large Hadron Collider~\cite{lhc}, yet no signs of SUSY have been observed so far. In view of these null results in \emph{conventional} SUSY searches, it becomes  mandatory to fully explore \emph{non-standard} SUSY scenarios involving $R$-parity violation (RPV)~\cite{rpv} and/or quasi-stable particles. 

$R$~parity is defined as $R = (-1)^{3(B-L)+2S}$, where $B$ ($L$) is the baryon (lepton) number and $S$ the spin, respectively. Hence all Standard Model particles have $R=+1$, whereas $R=-1$ for all SUSY particles. It is worth emphasising that the conservation of $R$~parity is an \emph{ad-hoc} assumption. The only strict limitation comes from the proton lifetime: non-conservation of both $B$ and $L$ leads to a rapid proton decay. $R$-parity conservation has serious consequences in SUSY phenomenology in colliders: the SUSY particles are produced in pairs and, most importantly, the lightest supersymmetric particle (LSP) is absolutely stable and weakly interacting, thus providing the characteristic high transverse missing momentum (\met) in SUSY events at colliders. Here we highlight two RPV models: the bilinear RPV (bRPV) and the $\mu$-from-$\nu$ supersymmetric standard model (\mn), which ---as we shall see--- both reproduce correctly the neutrino physics observations.

The structure of this paper is as follows. Section~\ref{sc:brpv} is dedicated to bRPV SUSY models: their connection with neutrino physics is explained in Section~\ref{sc:brpv-nu}, its phenomenology at the LHC is provided in Section~\ref{sc:brpv-pheno}, while Section~\ref{sc:brpv-atlas} presents the current constraints from the ATLAS experiment. The $\mu$-from-$\nu$ supersymmetric standard model is discussed in Section~\ref{sc:munussm}. After a brief review of the theoretical motivation for \mn, descriptions of its possible signatures at the LHC based on multileptons and displaced vertices follow in Section~\ref{sc:munussm-lhc}. Sections~\ref{sc:munussm-higgs} and ~\ref{sc:munussm-wz} focus on Higgs decays and unusual $Z$/$W$ boson decays in the context of $\mu\nu SSM$, respectively. Aspects of RPV SUSY linked to dark matter are discussed in Section~\ref{sc:dm}. The paper concludes with a summary and an outlook in Section~\ref{sc:summary}.

\section{Bilinear $R$-parity breaking}\label{sc:brpv}

$R$-parity conservation (RPC) has several consequences such as the stability of the LSP, which is a weakly interacting massive particle (WIMP) and consequently a candidate for dark matter (DM)~\cite{dm-review}. Being a WIMP, once produced at the LHC, it will escape detection, resulting in large missing transverse momentum, \met. Providing a DM candidate is one of the strongest arguments in favour of RPC SUSY, nonetheless RPV models \emph{do} exist that can explain DM through, for instance, very light gravitinos~\cite{gravitino,martin,brpv-dm,brpv-split,grefe,steffen,trpv-gravitino}, axions~\cite{axion,steffen} or axinos~\cite{axino,steffen,brpv-axino}. There is no fundamental reason for $R$~parity to be conserved thus lepton and baryon number violating renormalisable terms may appear in the supersymmetric potential in the following way:
\begin{equation}  
W= \lambda_{ijk}   L_i  L_j  E_k^c 
+\lambda_{ijk}'    L_i  Q_j  D_k^c 
+\epsilon_i        L_i  H_u
+\lambda_{ijk}''   U_i^c  D_j^c  D_k^c,
\label{eq:Wsuppot} 
\end{equation} 

\noindent where the couplings $\lambda$, $\lambda'$ and  $\lambda''$  are $3\times 3$ Yukawa matrices ---$i, j, k$ being flavour indexes---, $\epsilon_i$ are parameters with units of mass and $ Q, U, D, L, E, H_u, H_d $ refer to supermultiplets. The first three types of terms lead to lepton number violation, while the baryon number is violated by the fourth one.

As long as the breaking of $R$~parity is spontaneous, only bilinear terms arise in the effective theory below the RPV scale, thus rendering bilinear $R$-parity violation a theoretically attractive scenario. Moreover, the bilinear model provides a theoretically self-consistent scheme in the sense that trilinear RPV implies, by renormalisation group effects, that also bilinear RPV is present, but not conversely~\cite{Porod:2000hv}. 
 
In other words, the simplest way to break $R$~parity is to add bilinear terms to the MSSM potential. Besides that, additional bilinear soft SUSY breaking terms are introduced, which include small vacuum expectation values for the sneutrinos. The relevant superpotential $W$ and the soft supersymmetry breaking terms $V_{\rm soft}$, which include bilinear $R$-parity violation, would then be~\cite{Romao:1999up}
\begin{eqnarray}
\ W &=&  W^{\rm MSSM} + \epsilon_i  \hat{L}_i  \hat{H}_u \\
 V_{\rm soft} &=& V_{\rm soft}^{\rm MSSM}- B_i\epsilon_i  \tilde{L}_i  H_u\, ,
\end{eqnarray} 

\noindent where the $B_i$ have units of mass. In fact, if SUSY was not broken, the bilinear terms could be rotated away and be converted into trilinear terms, however the presence of the soft SUSY breaking terms $B_i\epsilon_i  \tilde{L}_i  H_u$ gives bRPV a physical meaning~\cite{Romao:1999up}. 

There are nine new parameters introduced in this model:  $\epsilon_i$, $B_i$ and $v_i$, the latter being the sneutrino vacuum expectation values (VEVs). However, after electroweak symmetry breaking and taking into account constraints from neutrino experiment results, only one free parameter remains in the model, which is set to be of the same order as the others. 

\subsection{Connection with neutrino physics}\label{sc:brpv-nu}

Sneutrino VEVs introduce a mixing between neutrinos and neutralinos, leading to a see-saw mechanism that gives mass to one neutrino mass scale at tree level. The second neutrino mass scale is induced by loop effects~\cite{Hirsch:2000ef,Diaz:2003as}. The same VEVs are also involved in the decay of the LSP, which in this case is the lightest neutralino. This implies a relation between neutrino physics and some features of the LSP modes. An example of such a connection is given by the relation~\cite{Hirsch:2000ef}
\begin{equation}
\label{tetatm}
\tan^2\theta_{\rm atm} = \left|{\frac{\Lambda_{\mu}}{\Lambda_{\tau}}}\right|^2 
\simeq \frac{BR(\tilde\chi_1^0\to\mu^\pm W^\mp)}{BR(\tilde\chi_1^0\to\tau^\pm W^\mp)} 
= \frac{BR(\tilde\chi_1^0\to\mu^\pm q \bar{q}')}{BR(\tilde\chi_1^0\to\tau^\pm q \bar q')},
\end{equation}
 
\noindent where $\theta_{\rm atm}$ is the atmospheric neutrino mixing angle and the ``alignment'' parameters $\Lambda_i$ are defined as $\Lambda_i = \mu v_i + v_d \epsilon_i$, with $v_d$ the VEV of $H^d$. This relation between RPV and neutrino physics allows setting bounds on bRPV parameters from results of neutrino experiment and astrophysical data~\cite{Abada:2000xr}. In the opposite direction, a possible positive signal observed in colliders may lead to the determination of some of the bRPV phenomenological properties, which in turn can constrain neutrino-physics parameters~\cite{brpv-nu}.

\subsection{Phenomenology at the LHC}\label{sc:brpv-pheno}

In the specific bilinear $R$-parity violating model discussed here, the LSP is the lightest neutralino, $\tilde\chi_1^0$~\cite{deCampos:2007bn}. The bRPV terms are embedded in the minimal Supergravity (mSUGRA) model, which imposes some restrictions which reduce the large number of parameters of the MSSM. The number of parameters in mSUGRA is reduced to five, namely $m_0$, the scalar mass; $m_{1/2}$, the gaugino mass; $A_0$, the trilinear scalar coupling; $\tan \beta = v_u / v_d$, the ratio of the Higgs VEVs; and ${\rm sgn}\,\mu $, the sign of the higgsino mass parameter.

The six bRPV parameters for each mSUGRA point are determined by the \texttt{SPheno}~\cite{spheno} spectrum calculator. \texttt{SPheno} uses as input the mSUGRA parameters and the neutrino physics constraints, and delivers as output the bRPV parameters (together with the mass spectra and the decay modes) compatible with these constraints. Once those quantities are calculated for a given set of bRPV parameters within an mSUGRA benchmark point, they can subsequently passed to an event generator and produce collision events at the LHC. 

\begin{figure}[htb]
\begin{minipage}{13pc}
\includegraphics[width=\textwidth]{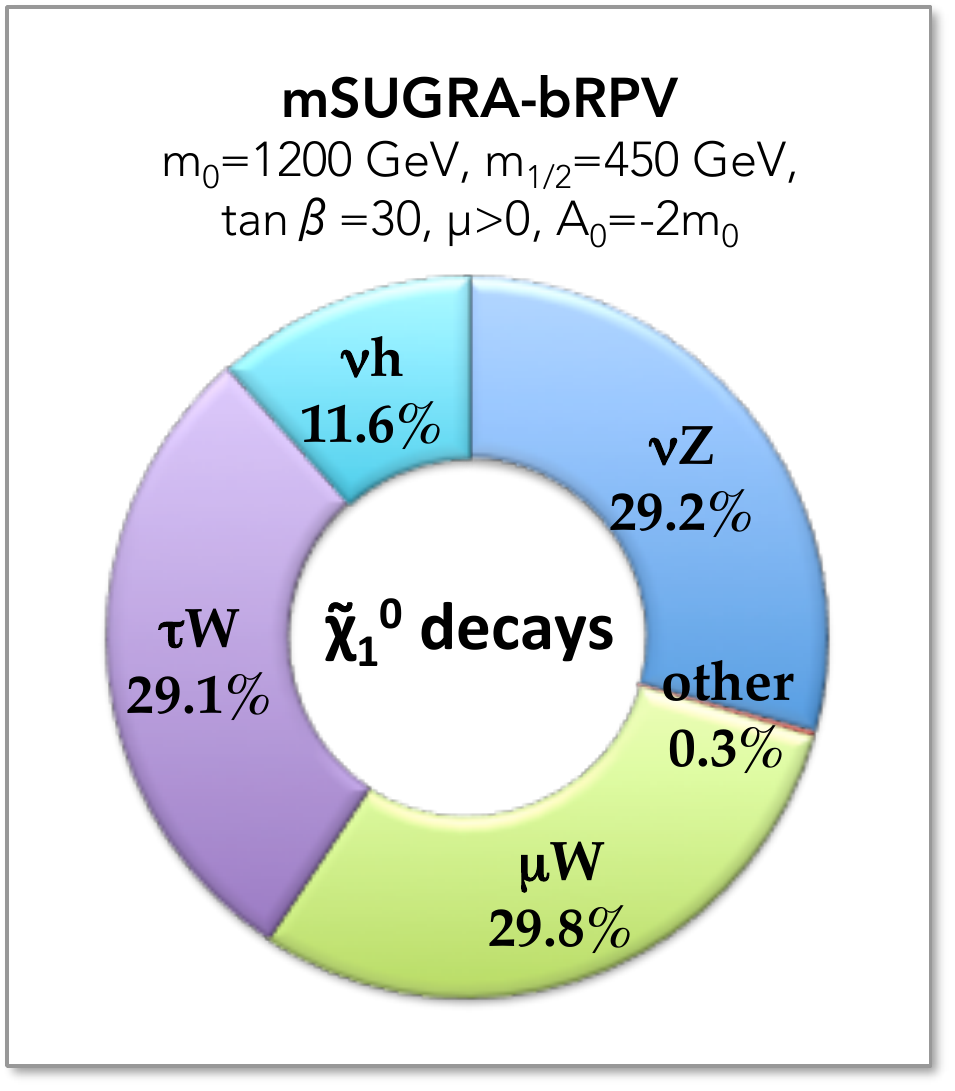}
\caption{\label{fg:decay}\X\ decay modes for a bRPV-mSUGRA point.}
\end{minipage}\hspace{2pc}%
\begin{minipage}{22pc}
\includegraphics[width=\textwidth]{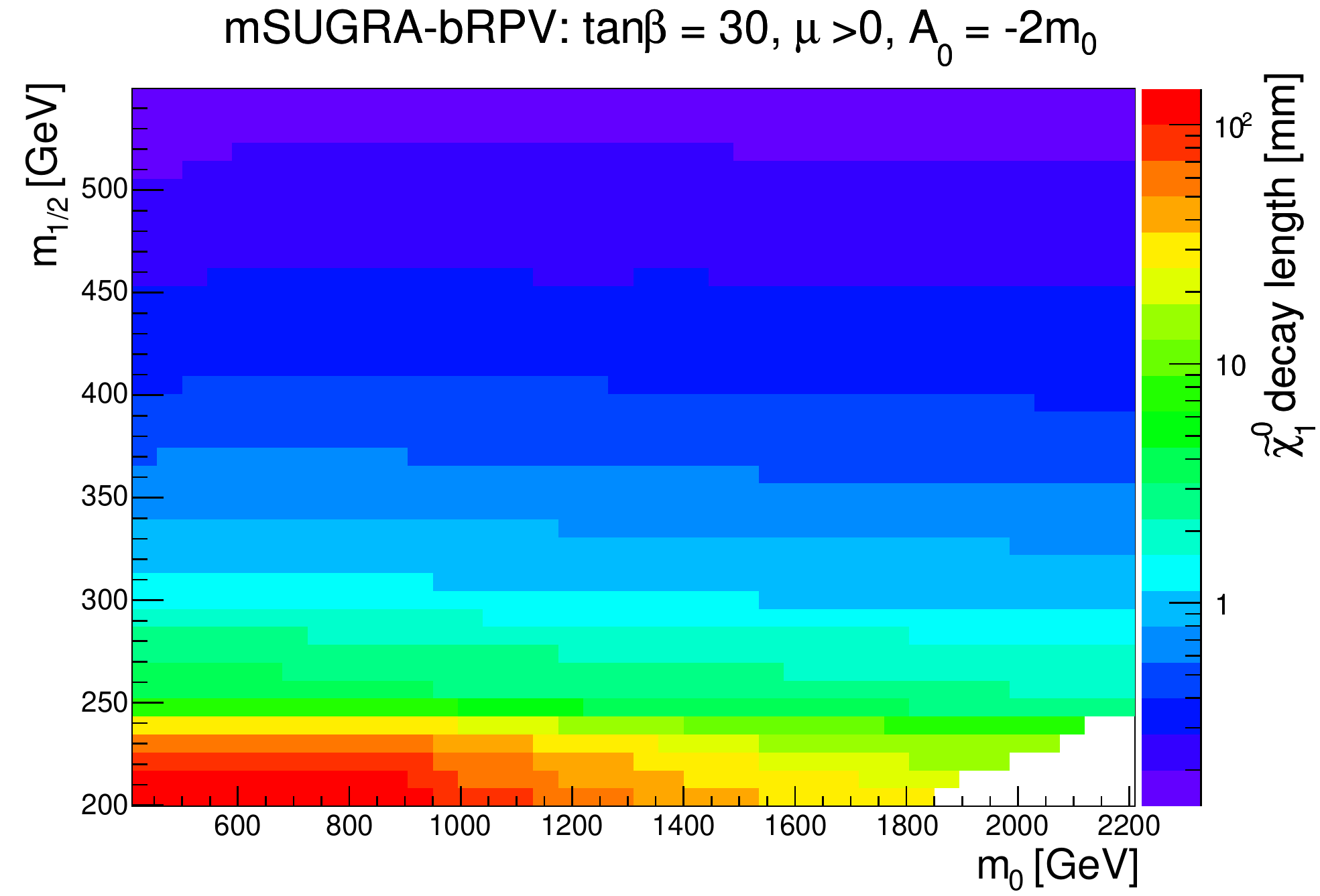}
\caption{\label{fg:decay-length}Proper decay length $c\tau$ in millimetres for the lightest neutralino LSP in bRPV-mSUGRA parameter plane $(m_0,\,m_{1/2})$ and $\tan\beta=30$, $\mu>0$ and $A_0 = -2m_0$.}
\end{minipage} 
\end{figure}

It is stressed that the sparticle spectrum for bRPV-mSUGRA is ---within theoretical uncertainties--- the same as in RPC mSUGRA; it is the LSP decay modes and its lifetime that depend on the bRPV parameters. Typical \X\ decay modes are shown in Fig.~\ref{fg:decay} for a model where the parameters $\tan\beta=30$, $\mu>0$ and $A_0=-2m_0$ have been chosen so as to obtain a \mbox{125-\gev} Higgs boson in agreement with the ATLAS~\cite{atlas-higgs} and CMS~\cite{cms-higgs} discovery. The neutralino decays are dominated by leptonic $(e,\,\mu)$ or $\tau$ channels, making lepton-based searches ideal for constraining this model. The decay to a muon and a hadronically-decaying $W$ has been studied in detail and it has been shown that it may lead to the measurement of the \X\ mass through the reconstruction of the $\X\to\mu q'\bar{q}$ peak in the $\mu jj$ invariant mass~\cite{int-note,emma,brpv-nu}. Furthermore the fact that in the low-$m_{1/2}$ high-$m_0$ region the \X\ is slightly long lived, as evident from Fig.~\ref{fg:decay-length}, opens up the possibility to use searches for displaced vertices~\cite{atlas-dv,atlas-dv-dilep} in order to probe this model at the LHC~\cite{brpv-dv}. Lastly, the \X\ decays to one or two neutrinos give rise to a moderately high \met, thus rendering some \met-based analyses pertinent to bRPV, as we shall see in the next section.

So far we have discussed the bRPV in the context of supergravity models. If the bRPV terms are introduced to anomaly-mediated SUSY breaking (AMSB) instead, the \X\ LSP decay modes do not change, whereas the wino-like neutralino (as opposed to the bino-like in mSUGRA) is characterised by stronger interactions, hence it is easier to be produced at LHC. The mass degeneracy of the \X\ and the $\t{\chi}^{\pm}_1$ makes the $\t{\chi}^{\pm}_1$ long lived, decaying dominantly through RPV couplings to $\ell\ell\ell$, $\tau\ell\ell$, $\ell b\bar{b}$, $\tau b\bar{b}$~\cite{brpv-amsb}. In other words, the common feature of bRPV phenomenology is the presence of delayed decays leading to interesting signatures at the LHC.

\subsection{Constraints from ATLAS}\label{sc:brpv-atlas}

As mentioned above, thanks to the abundant neutrino production from the LSP decay, bRPV events at the LHC are expected to feature relatively (when compared to Standard Model processes) high missing transverse momentum. High lepton/$\tau$ multiplicity is also expected from the LSP decays and from upstream lepton production in the SUSY cascade decay, if strong production is considered. Both features are exploited when looking for a signal of the bRPV-mSUGRA model, as demonstrated in Ref.~\cite{int-note}, where a first detailed Monte-Carlo-based study on the observability of the $\X\to\mu q'\bar{q}$ at the LHC has been carried out. The very first bounds set on a bilinear RPV model in colliders were provided by an inclusive search for high \met, three or four jets and one muon at $\sqrt{s}=7~\tev$ and $\sim 1~\ifb$ of ATLAS data~\cite{brpv-atlas1,emma}. The \X\ decay to muonic channels is enhanced when compared to electrons, as shown in Fig.~\ref{fg:decay}, hence muon channels yield stronger limits than their electron counterparts. A slight loss of sensitivity in the high-$m_0$ low-$m_{1/2}$ region was observed due to the large LSP lifetime, which causes signal muons to be rejected by the cosmic-muon veto, i.e.\ the cut on muons with high impact parameter. Aforesaid limits were extended further by an analysis with $5~\ifb$ based on events with high jet multiplicity (at least seven jets), large \met\ and exactly one lepton (muon or electron)~\cite{brpv-atlas2,vam}.

Further constraints on bRPV-SUGRA for parameters tuned to attain a mass of the lightest Higgs equal to 125~\gev\ have been set recently by ATLAS~\cite{atlas-ss-lep,atlas-tau,atlas-leptons} with the full data set of $\sim 20~\ifb$ recorded at $\sqrt{s}=7~\tev$. The LSP decays to taus have been used as a handle to constrain the model in an analysis combining event selections with at least one tau, two taus or one tau and a lepton ($e$ or $\mu$) in conjunction with large \met~\cite{atlas-tau}. The observed limits can be consulted in the left panel of Fig.~\ref{fg:brpv-atlas}. The bounds set by an analysis seeking two same-sign leptons or three $b$-jets and jets~\cite{atlas-ss-lep} are significantly extended with respect to the $\tau$-analysis. As shown in the right panel of Fig.~\ref{fg:brpv-atlas}, values of $m_{1/2}$ are excluded between 200~\gev\ and 490~\gev\ at 95\% CL for $m_0<2.1~\tev$. Interestingly enough, in both analyses, signal points with $m_{1/2}<200~\gev$ were not considered because the lepton acceptance was significantly reduced due to the increased LSP lifetime in that region, as seen previously in Fig.~\ref{fg:decay-length}. Slightly stringent limits have been set by ATLAS very recently by requiring one hard lepton, jets and large \met~\cite{atlas-leptons}. 

The fact that in a large part of the parameter space the \X\ LSP decays to a $Z$~boson and a neutrino implies that searches for events containing $Z$ and large \met~\cite{cms-z,atlas-z} should be pertinent to explore bRPV as well. In fact this LSP decay ---as much as decays to $\tau W$ and $\mu W$--- is considerable not only when bRPV is embedded in mSUGRA, but also when bRPV is introduced to other well-motivated SUSY scenarios such as ``natural'' SUSY~\cite{nat-susy} and pMSSM~\cite{pmssm}. Ongoing work continues in interpreting the results of the same-sign lepton analysis~\cite{atlas-ss-lep} in natural pMSSM~\cite{nat-pmssm} with bRPV~\cite{atlas-rpv-summary}.
 
\begin{figure}[htb]
\centering
\includegraphics[width=0.485\textwidth]{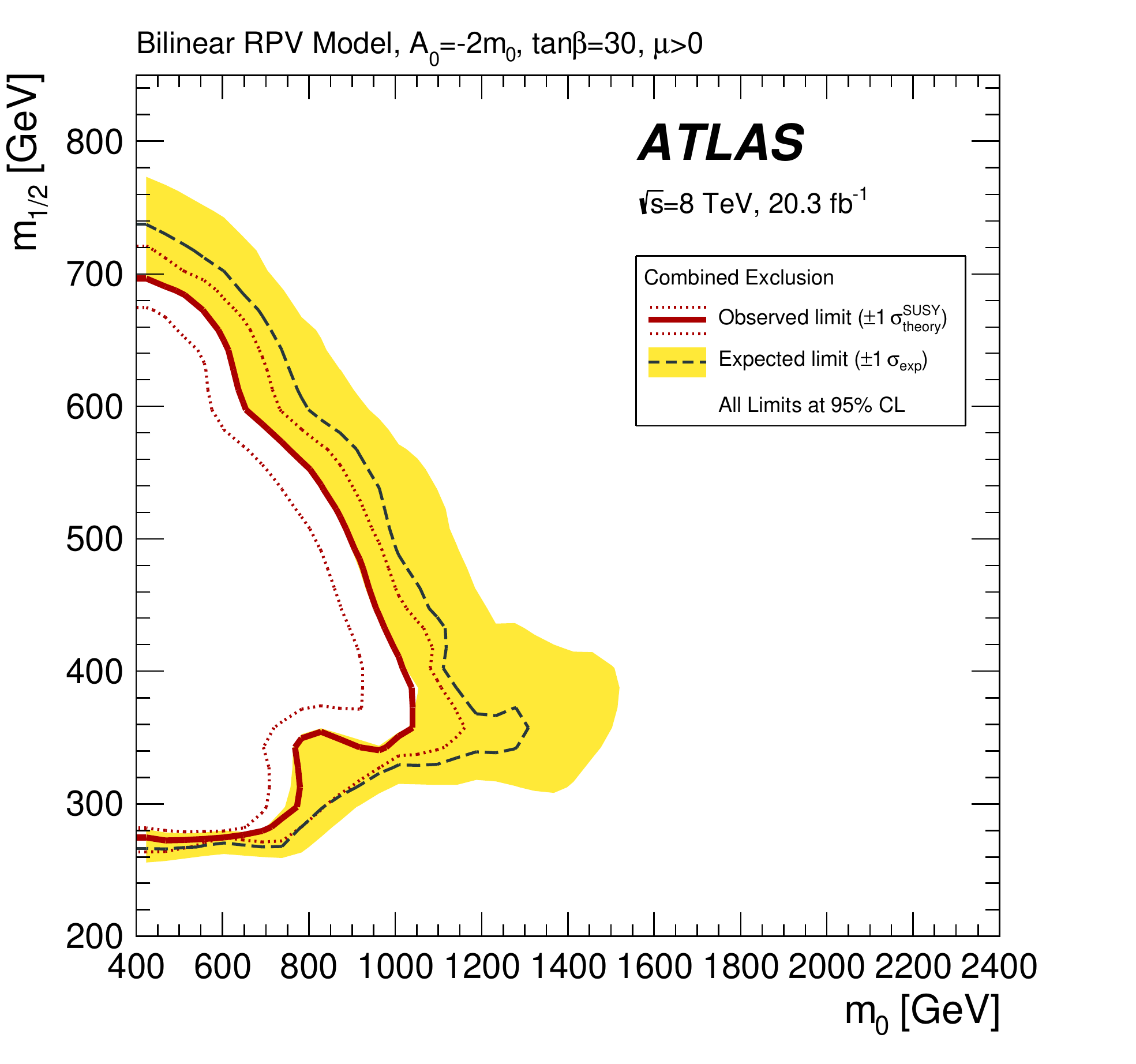}
\hfill
\includegraphics[width=0.475\textwidth]{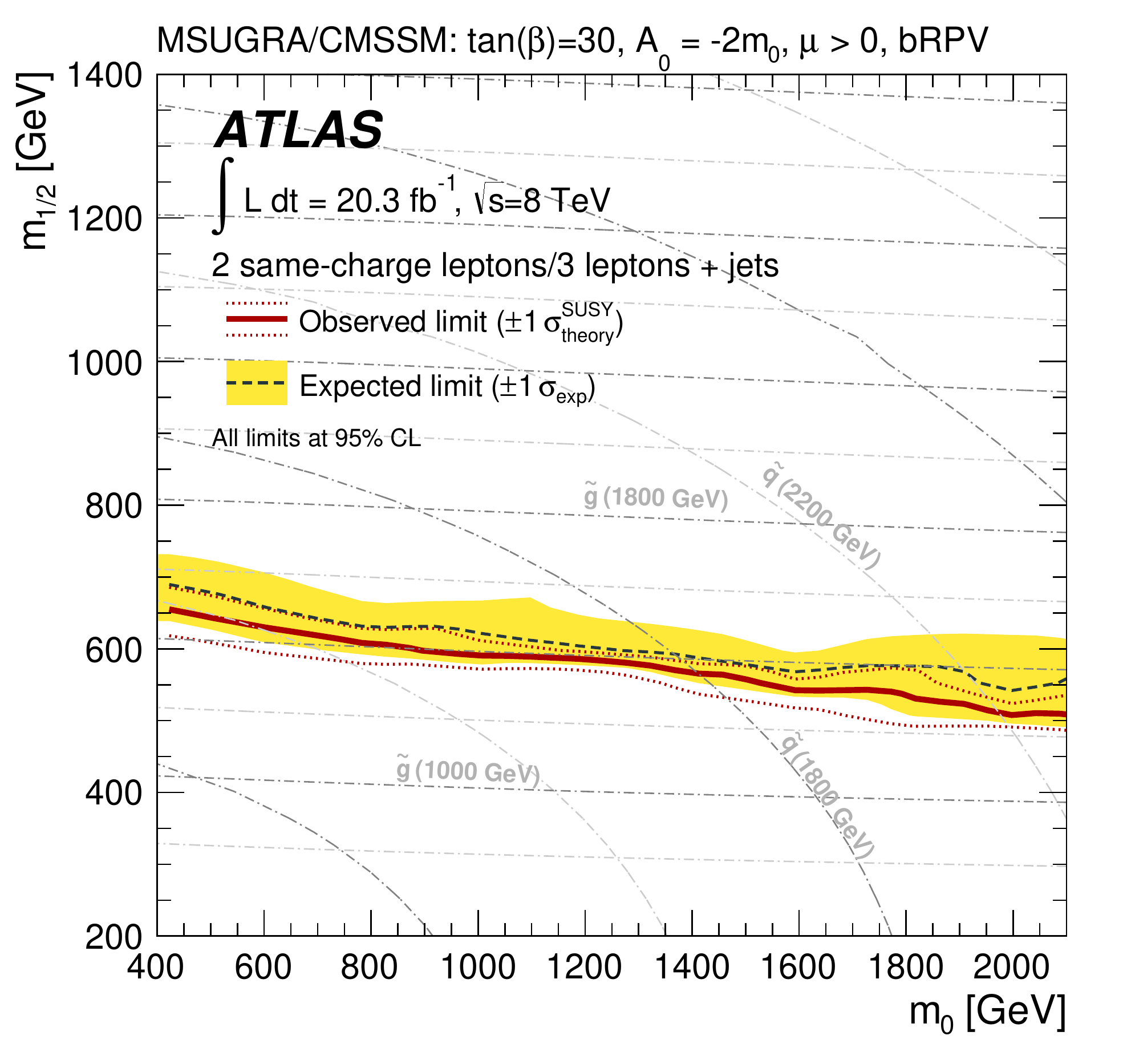}
\caption{Expected and observed 95\% CL exclusion limits in the bilinear $R$-parity violating model obtained by a $\tau+\met$ analysis (left)~\cite{atlas-tau} and by the two-same-sign-lepton analysis (right)~\cite{atlas-ss-lep}. The latter results are obtained by combining the electron and muon channels. The band around the median expected limit shows the $\pm 1\sigma$ variations on the median expected limit, including all uncertainties except theoretical uncertainties on the signal. The dotted lines around the observed limit indicate the sensitivity to $\pm 1\sigma$  variations on these theoretical uncertainties. } \label{fg:brpv-atlas} 
\end{figure}

\section{The $\mu$-from-$\nu$ supersymmetric standard model ($\mu\nu$SSM)}\label{sc:munussm}

The \mn~\cite{mnssm-proposal,mnssm-spectrum} is a supersymmetric standard model that solves the $\mu$~problem~\cite{mu-problem} of the minimal supersymmetric standard model (MSSM) using the $R$-parity breaking couplings between the right-handed neutrino superfields and the Higgs bosons in the superpotential, $\lambda_i\hat{\nu}_i^c\hat{H}_d\hat{H}_u$. The $\mu$~term is generated spontaneously through sneutrino vacuum expectation values, $\mu = \lambda_i\langle\t{\nu}_i^c\rangle$, once the electroweak symmetry is broken, without introducing an extra singlet superfield as in the case of the next-to-MSSM (NMSSM)~\cite{nmssm}. The complete \mn\ superpotential is given by
%
\begin{equation}\label{eq:w}
\begin{aligned}
W  = &
\ \epsilon_{ab} (
Y_{u_{ij}} \, \hat H_u^b\, \hat Q^a_i \, \hat u_j^c +
Y_{d_{ij}} \, \hat H_d^a\, \hat Q^b_i \, \hat d_j^c +
Y_{e_{ij}} \, \hat H_d^a\, \hat L^b_i \, \hat e_j^c 
\\ 
&
+ Y_{\nu_{ij}} \, \hat H_u^b\, \hat L^a_i \, \hat \nu^c_j -   
\lambda_{i} \, \hat \nu^c_i\,\hat H_d^a \hat H_u^b)+
\frac{1}{3}
\kappa{_{ijk}} 
\hat \nu^c_i\hat \nu^c_j\hat \nu^c_k.
\end{aligned}
\end{equation}

The couplings $\kappa_{ijk}\hat{\nu}_i^c\hat{\nu}_j^c\hat{\nu}_k^c$ forbid a global $U(1)$ symmetry avoiding the existence of a Goldstone boson, and also contribute to spontaneously generate Majorana masses for neutrinos at the electroweak scale. The latter feature is unlike the bilinear RPV model, where, as mentioned in Section~\ref{sc:brpv-nu}, only one mass is generated at the tree level and loop corrections are necessary to generate at least a second mass and a neutrino mixing matrix compatible with experiments. In the bilinear model, the $\mu$-like problem~\cite{mu-problem-nilles} is augmented by the presence of three bilinear terms.

The \mn\ phenomenology is largely defined by the parameters
\begin{equation}
\boldsymbol{\lambda},\:\kappa_{i}, \:\nu^c,\: \tan\beta, \:M_1, \:A_{\lambda}, \: A_\kappa\, ,
\label{EWF-param3A}
\end{equation}
where $\boldsymbol{\lambda}\equiv\sqrt{3}\lambda$ it the singlet-doublet mixing parameter (if universal $\lambda_i$ are assumed), $\kappa$ is the  common $\kappa_{ijk}$, $A_{\lambda}, A_{\kappa}$ are the soft SUSY-breaking parameters and $M_1$ is the $U(1)$ mass scale.

In the \mn, as a consequence of $R$-parity violation, all neutral fermions mix together into ten neutralinos, $\tilde{\chi}_{\alpha}^0$, and five charginos, $\tilde{\chi}_{\alpha}^{\pm}$. Since the three lightest neutralinos are the left-handed neutrinos, the ``true'' LSP would be the $\tilde{\chi}_4^0$. Likewise the three lightest charginos $\tilde\chi^\pm_i$ coincide with the three charged leptons. Similarly, all scalars mix into eight $CP$-even, $S_{\alpha}^0$ and seven $CP$-odd neutral Higgs bosons, $P_{\alpha}^0$,  mass eigenstates. The three lighter neutral scalars, $S_i^0, i=1, 2, 3$, are in fact naturally light singlet-like states, leaving the fourth one, $S_4^0$, to play the role of the discovered SM-Higgs-like scalar. Charged scalars, on the other hand, form seven mass eigenstates, $P_{\alpha}^{\pm}$. Analyses of the \mn, with attention to the neutrino and LHC phenomenology have been addressed in Refs.~\cite{mnssm-spectrum,mnssm-others}. Other analyses concerning cosmology such as gravitino dark matter and electroweak baryogenesis can be found in Refs.~\cite{mnssm-gravitino1,mnssm-gravitino2,mnssm-fermi} and in Ref.~\cite{mnssm-baryo}, respectively. In conclusion, \mn\ is a well-motivated SUSY model with enriched phenomenology and notable signatures, which definitely deserve rigorous analyses by the LHC experiments. Its enlarged Higgs sector can easily accommodate the observed 125~\gev\ Higgs boson~\cite{atlas-higgs,cms-higgs}. 

\subsection{Signatures at LHC}\label{sc:munussm-lhc}

Here a collider analysis together with detector simulation of an intriguing signal in the \mn\ featuring non-prompt multileptons at the LHC, arising from the beyond SM decay of a 125~\gev\ scalar into a pair of lightest neutralinos, \XX, is presented~\cite{my-mnssm1}. Since $R$~parity is broken, each \XX\ decays into a scalar/pseudoscalar $(h/P)$ and a neutrino, with the $h/P$ further driven to decay into $\tau^+ \tau^-$, giving rise to a $4\tau$ final state, as shown in Fig.~\ref{fg:decay-chain}. The small $R$-parity breaking coupling of \XX\ renders it long~lived, yet it decays inside the inner tracker, thereby yielding clean detectable signatures: (i) high lepton multiplicity; and (ii) charged tracks originating from displaced vertices (DVs). 

\begin{figure}[htb]
\includegraphics[width=15pc]{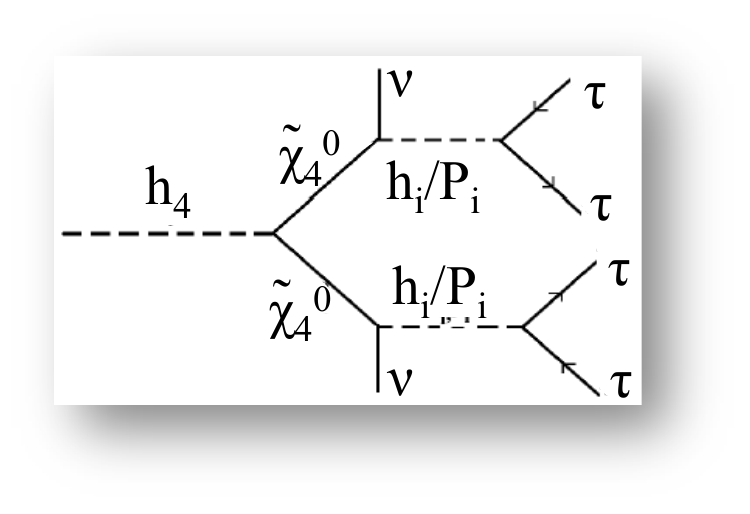}\hspace{2pc}%
\begin{minipage}[b]{20pc}\caption{\label{fg:decay-chain}The \mn\ decay chain studied in Ref.~\cite{my-mnssm1}. The Higgs boson is produced through gluon fusion and has a mass of 125~\gev. The neutralino \XX\ is long-lived and gives rise to displaced $\tau$~leptons.}
\vspace*{0.83cm}
\end{minipage}
\end{figure}

The \mn\ is characterised by the production of several high-\pt\ leptons~\cite{my-mnssm1}. Electrons and muons are produced through the leptonic $\tau$ decays, yet muon pairs can appear directly through $h_i/P_i$ decays as well. With the chosen decay mode, the $\tau$ multiplicity is considerable even though the $\tau$-identification efficiency is much lower ($\sim50\%$) when compared to that of electrons and muons ($\gtrsim95\%$). Occasionally highly collimated QCD jets can fake hadronic $\tau$~leptons, $\tau_\text{had}$, and, as a result, $\tau$ multiplicity may exceed the expected number of four. This faking however disappears with a higher \pt~cut, which should also be sufficient to provide a single-lepton trigger for the analysis. 

\begin{figure}[htb]
\begin{minipage}{16.5pc}\vspace*{0.2cm}
\includegraphics[width=\textwidth]{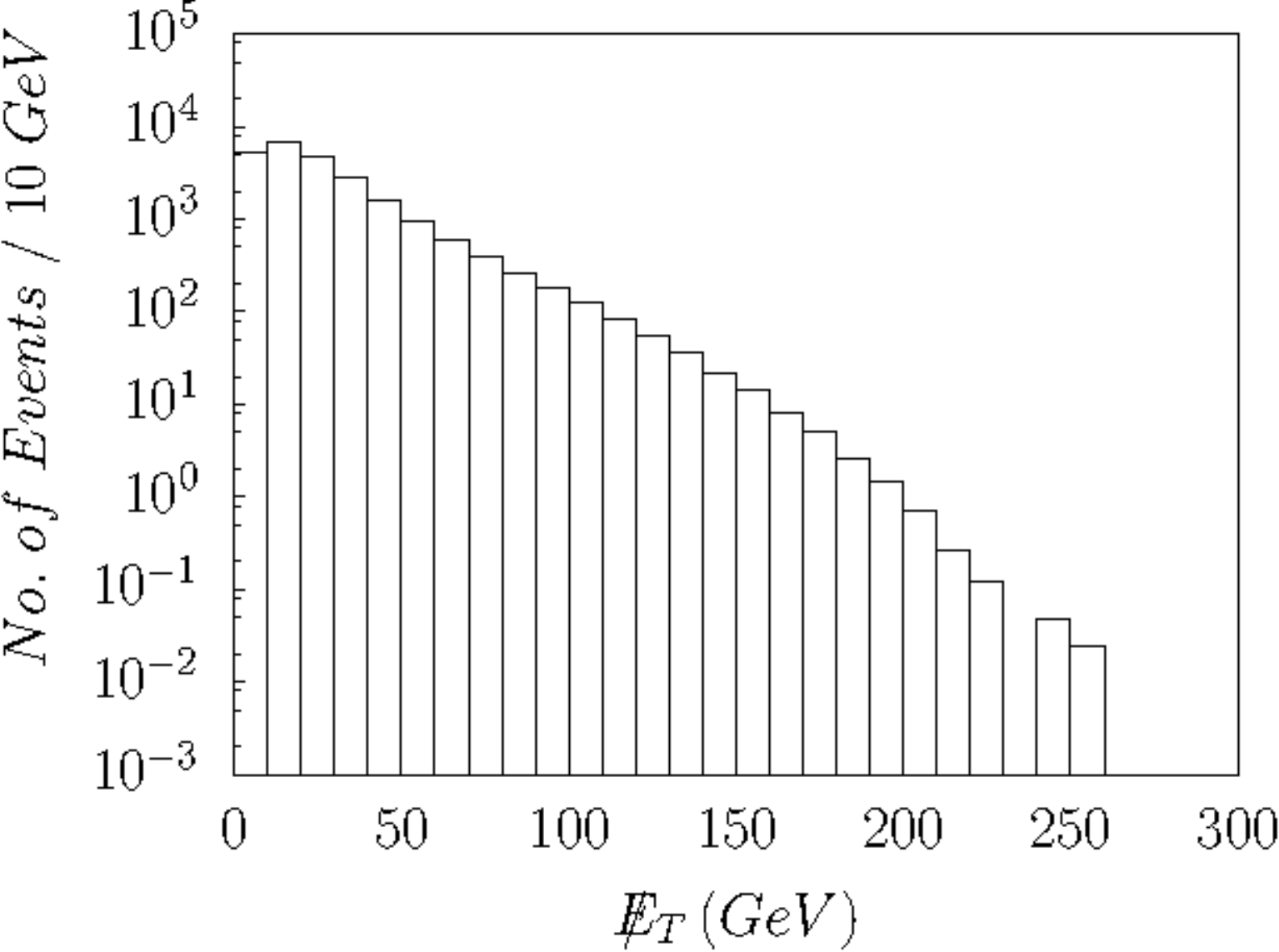}
\caption{\label{fg:met} \met~distribution for an LHC energy of $\sqrt{s}=8~\tev$ and an integrated luminosity $\mathcal{L}=20~\ifb$ for a selected \mn\ point~\cite{my-mnssm1}.}
\end{minipage}\hspace{2pc}%
\begin{minipage}{19pc}
\includegraphics[width=\textwidth]{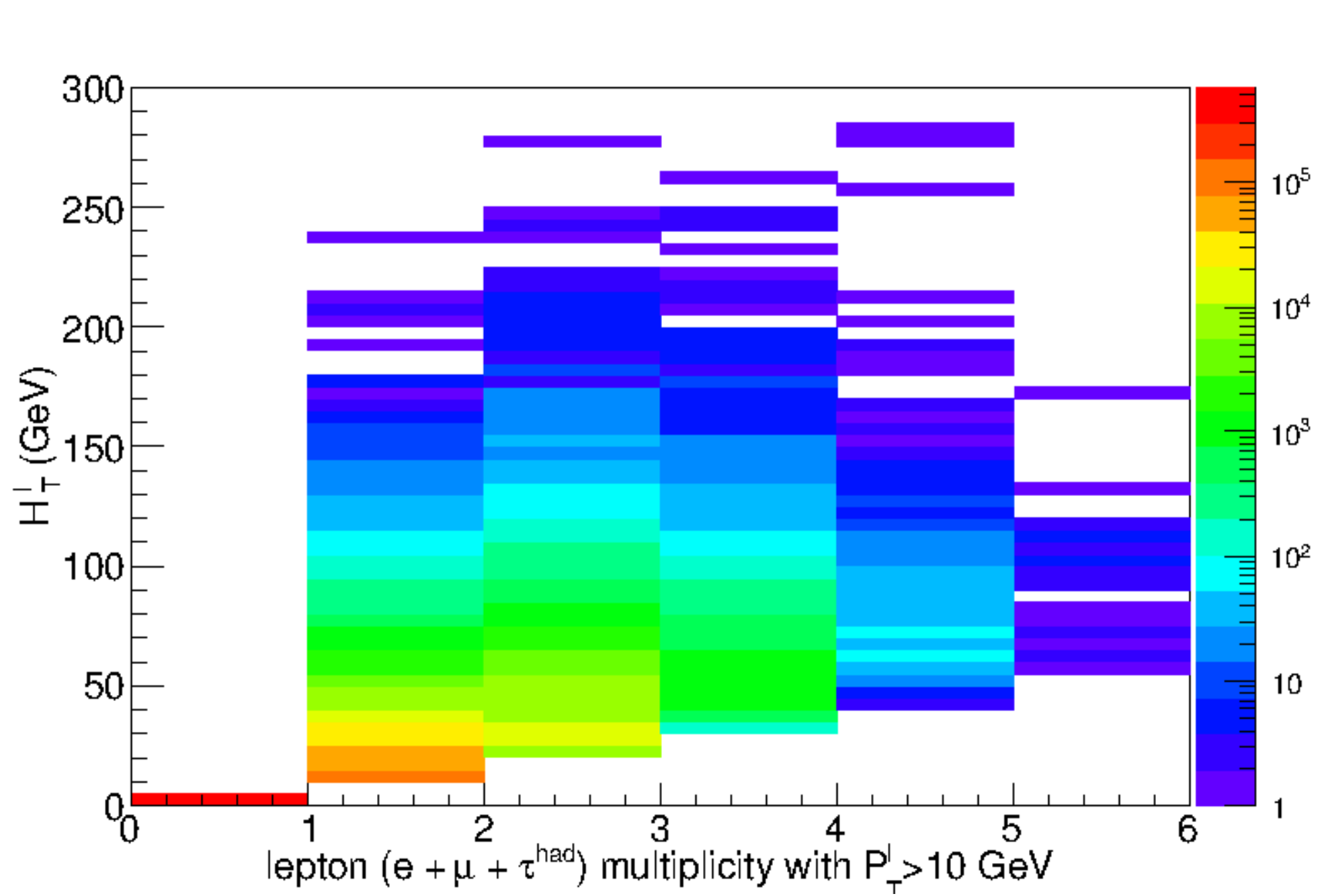}
\caption{\label{fg:HTvslnhigh-cut}$H{\rm _T^\ell}$ versus lepton multiplicity for an LHC energy of $\sqrt{s}=8~\tev$ and an integrated luminosity $\mathcal{L}=20~\ifb$ for a selected \mn\ point~\cite{my-mnssm1}.}
\end{minipage} 
\end{figure}

Apart from the requirement of at least three or four leptons (including taus), a high value of \met\ and/or of the scalar sum of reconstructed objects: leptons, jets and/or \met\ is needed~\cite{multileptons}. For the chosen signal many neutrinos ($\geq 6$) appearing in the final state from \XX\ and from $\tau$ decay give rise to moderately high ---when compared to signals from RPC supersymmetry--- missing transverse energy, \met, as depicted in Fig.~\ref{fg:met}. Besides \met, the scalar sum of the \pt\ of all reconstructed leptons, $H{\rm _T^\ell}$, is also large in such events, as shown in Fig.~\ref{fg:HTvslnhigh-cut}. Alternatively, the sum of \met\ and $H{\rm _T^\ell}$ can be deployed for further background rejection. These observables can provide additional handles when selecting events with many leptons. In addition, the invariant masses, $m_{\ell^+\ell^-}$ and $m_{2\ell^+2\ell^-}$ may prove useful for the purpose of signal distinction. 

\begin{figure}[htb]
\begin{minipage}{16.5pc}
\includegraphics[width=\textwidth]{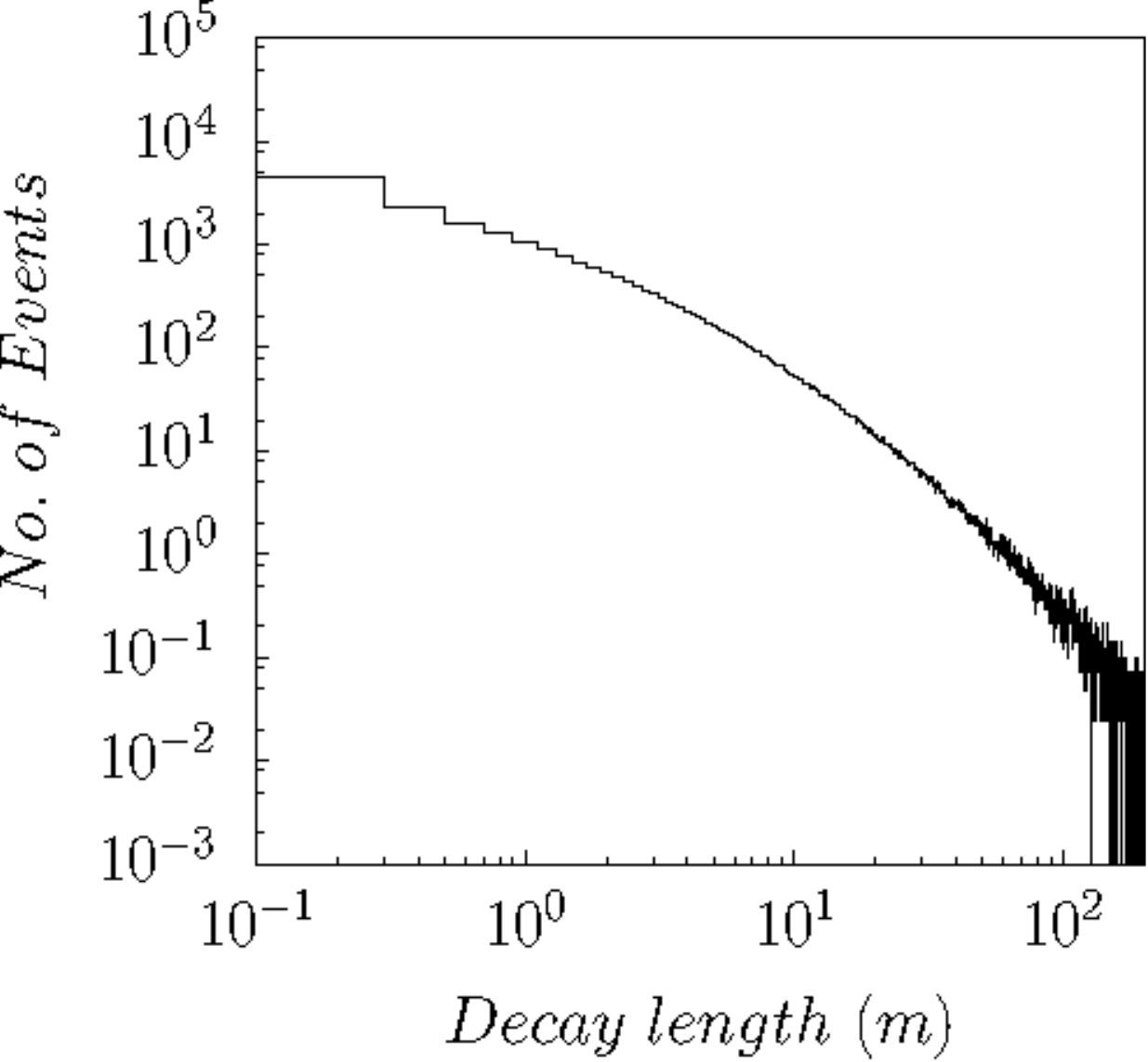}
\caption{\label{fg:dldistr} \XX\ decay-length distribution for an LHC energy of $\sqrt{s}=8~\tev$ and an integrated luminosity $\mathcal{L}=20~\ifb$ for a selected \mn\ point~\cite{my-mnssm1}.}
\end{minipage}\hspace{2pc}%
\begin{minipage}{19pc}
\includegraphics[width=\textwidth]{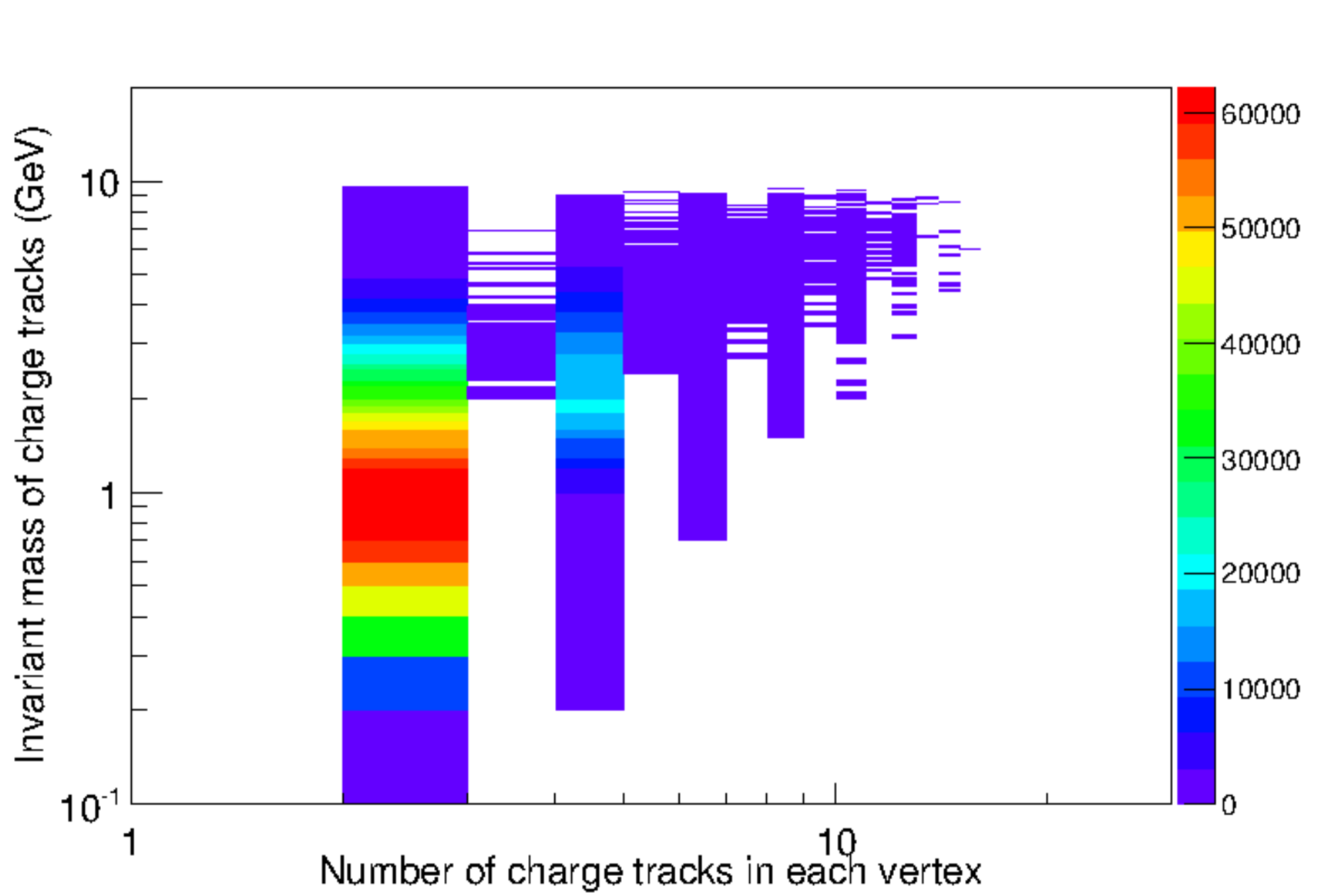}
\caption{\label{fg:DVtrackvsmass12C}Charged-track mass versus the number of charge particles in each vertex for an LHC energy of $\sqrt{s}=8~\tev$ and an integrated luminosity $\mathcal{L}=20~\ifb$ for a selected \mn\ point~\cite{my-mnssm1}.}
\end{minipage} 
\end{figure}

In the benchmark scenario under study, the \XX\ is characterised by a proper lifetime of the order of  $\tau_{\XX}\approx10^{-9}$~s, which corresponds to a proper decay length of $c\tau_{\XX}\approx 30$~cm. This feature ---quantified in Fig.~\ref{fg:dldistr}, were the decay-length distribution is plotted--- gives rise to displaced vertices. In a significant percentage of events, the \XX\ would decay inside the inner tracker of the LHC experiments, e.g.\ in 28\% of events it would decay within $30~{\rm cm}$ and in $44$\% events within 1~m. Therefore, the \mn\ signal events would be characterised by displaced $\tau$~leptons plus neutrinos. This distinctive signature opens up the possibility to exploit current or future variations of analyses carried out by ATLAS and CMS looking for a displaced muon and tracks~\cite{atlas-dv,atlas-dv-dilep} or searching for displaced dileptons~\cite{cms-dileptons,atlas-dv-dilep}, dijets~\cite{dijets} or  muon jets~\cite{atlas-dl} arising in Higgs decays to pairs of long-lived invisible particles. 

The kinematics of the DVs and their products in the chosen \mn\ benchmark point have been studied thoroughly in Ref.~\cite{my-mnssm1}. The spacial distribution of DVs shows that an appreciable fraction of them falls in the inner-tracker volume of the LHC experiments, i.e.\ $\rho_{\text{DV}}\lesssim 1$~m and $|z_{\text{DV}}|\lesssim 2.5$~m, thus DVs arising in the \mn\ should be detectable at LHC, in principle, either with existing analyses~\cite{atlas-dv,atlas-dv-dilep,cms-dileptons,dijets,atlas-dl} or via variations of those looking for displaced taus and \met. The average \XX\ boost, on the other hand, as expressed by $\beta\gamma$, where $\beta$ is \XX\ velocity over $c$ and $\gamma$ the Lorentz factor, is comparable to the signal analysed in an ATLAS related search for a muon and tracks originating from DVs~\cite{atlas-dv}. The boost affects the efficiency with which such a DV can be reconstructed, since high \XX\ boost leads to collimated tracks difficult to differentiate from primary vertices. In Fig.~\ref{fg:DVtrackvsmass12C}, the correlation between the number of charged tracks in each DV, $N_{\rm trk}$, and their invariant mass, $m_{\rm DV}$, is shown. It has been demonstrated~\cite{atlas-dv} that a selection of high-$N_{\rm trk}$ and high-$m_{\rm DV}$ efficiently suppresses background from long-lived SM particles ($B$~mesons, kaons). The modulation observed in $N_{\rm trk}$ is due to the one-prong or three-prong hadronic $\tau$ decays. 

\subsection{Higgs decays}\label{sc:munussm-higgs}

As pointed out earlier, \mn\ can easily predict a Higgs boson of a mass compatible with the observed one~\cite{atlas-higgs,cms-higgs}. As becomes evident from Fig.~\ref{fg:higgs-mass} a 125-\gev\ Higgs boson, $S_4^0$, can be accommodated within a wide range of $\tan\beta$ and $\boldsymbol{\lambda}$ values~\cite{my-mnssm2}. 

\begin{figure}[htb]
\begin{minipage}{17pc}
\includegraphics[width=\textwidth]{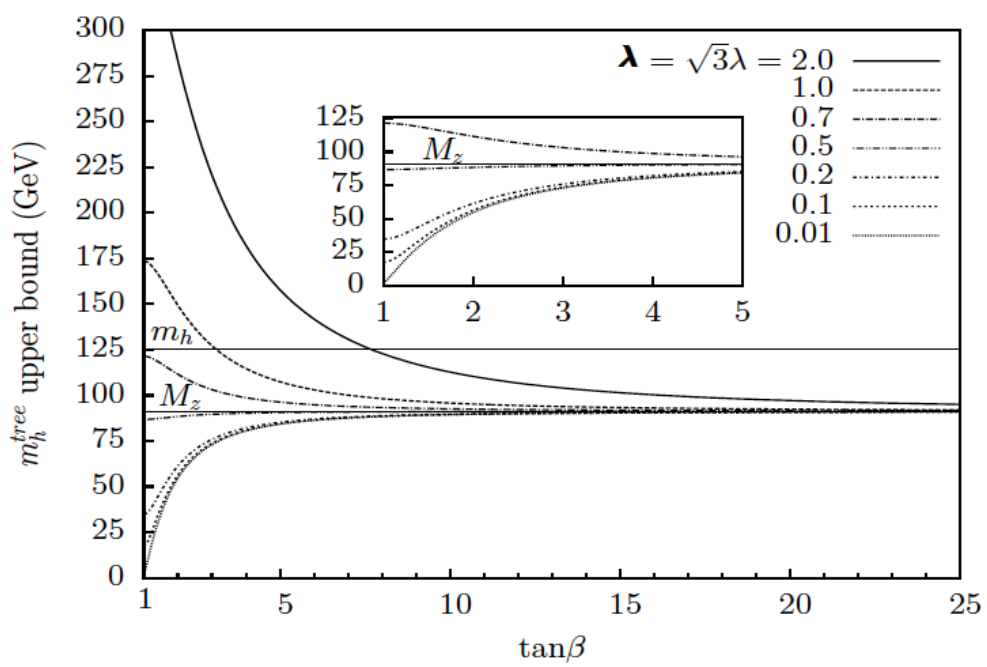}
\caption{\label{fg:higgs-mass} Variation of $m_h^{\rm tree}$ upper bound with $\tan\beta$ for different $\boldsymbol{\lambda}$ values in the \mn. The horizontal lines represent the experimentally measured masses of the Higgs, $m_h$, and $Z$~boson, $m_Z$. In the inset the region of $\tan\beta\leq5$ is shown for $\lambda\leq0.7$~\cite{my-mnssm2}.}
\end{minipage}\hspace{1.9pc}%
\begin{minipage}{19pc}
\includegraphics[width=\textwidth]{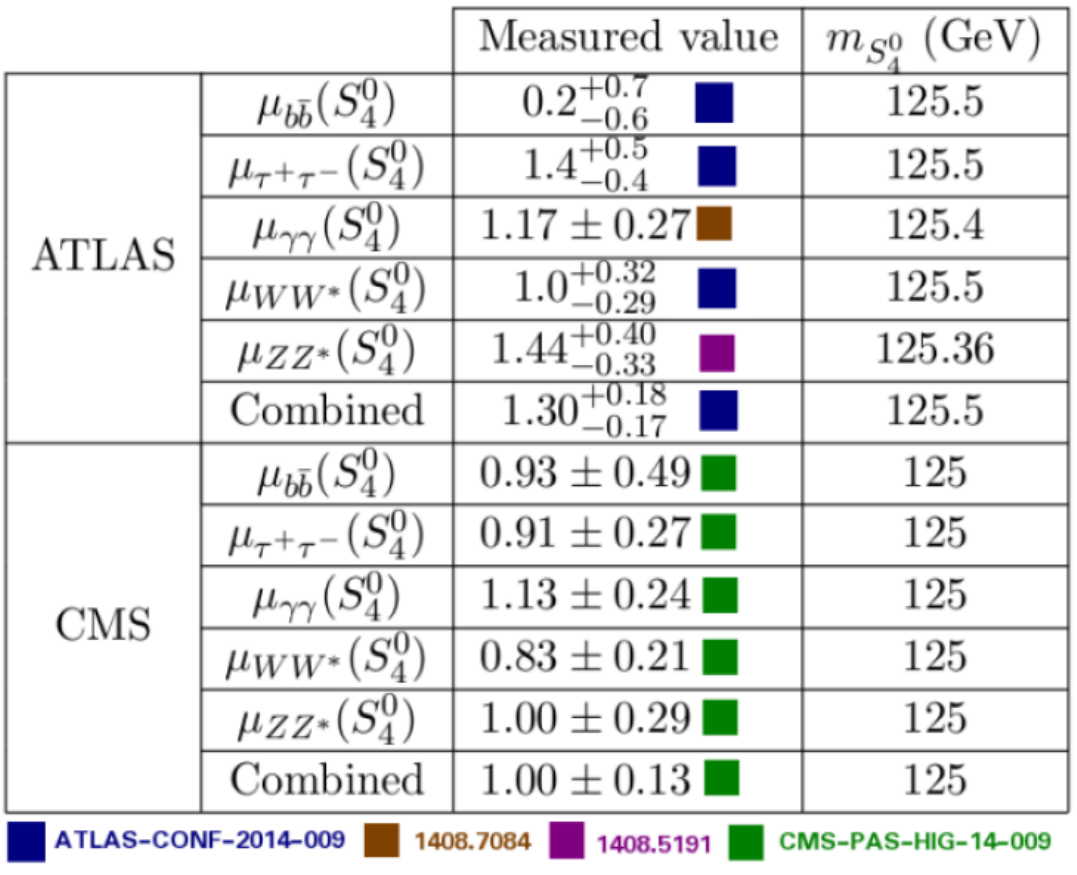}
\caption{\label{fg:higgs-signal} Measured signal strengths with $1\sigma$ uncertainties obtained from the ATLAS and CMS experiments together with the corresponding values of Higgs mass~\cite{my-mnssm2}.}
\end{minipage} 
\end{figure}

This scalar would produce new states through two-body decays as long as these states are kinematically allowed. Decays to $S_i^0S_j^0$, $P_i^0P_j^0$, $\tilde{\chi}_{i+3}^0\tilde{\chi}_{j+3}^0$ with $i,j = 1, 2, 3$ have been studied in detail in Ref.~\cite{my-mnssm2}. The question is whether these new decay widths respect the current measurements of the Higgs decays signal strength $\mu_{XX}$. It is shown that the final states are dominated by a combination of prompt or displaced leptons, taus, jets and photons plus \met\ due to neutrinos. When the expected productions rates for these new signatures are compared with the current $\mu_{XX}$ measurements listed in Fig.~\ref{fg:higgs-signal}, it is found that they are compatible with the measured Higgs signal strengths $\mu_{XX}$ under the following conditions.:
\begin{description}
\item[\underline{$0.01 < \boldsymbol{\lambda} < 0.1$}:] All $\mu_{XX}$ remain within $2\sigma$ of CMS measurements for $2.5 < \tan\beta < 3.9$.
\item[\underline{$0.1 < \boldsymbol{\lambda} < 0.7$}:] Only (invisible) $S_4^0$ decays to $\tilde{\chi}_{i+3}^0\tilde{\chi}_{j+3}^0$ remain viable in the whole range of $\boldsymbol{\lambda}$.
\item[\underline{$\boldsymbol{\lambda} > 0.1$}:] For decays to pair of binos, all $\mu_{XX}$ are within $2\sigma$ for $2.4 < \tan\beta < 3.8$.
\end{description}
Hence there is plenty of room for new Higgs decays within the context of the ``$\mu$ from $\nu$'' supersymmetric standard model. 

\subsection{New decays to $Z/W$}\label{sc:munussm-wz}

In addition to the new SM-Higgs-like decays, \mn\ also introduces novel on-shell decays of the $Z$ and $W^{\pm}$ gauge bosons~\cite{my-mnssm3}. These modes are typically encountered in regions of the parameter space populated with light singlet-like scalars, pseudoscalars and neutralinos.  The complete spectrum of possible final states and their origin is presented in Table~\ref{tb:wz}. The delayed ``objects'' occur in delayed decays of the neutralino, whereas the (short-lived) scalars and pseudoscalars deliver prompt products. 

\begin{table}[htb]
\caption{\label{tb:wz} Final states from non-standard $Z$ and $W^{\pm}$ decays with their respective origins~\cite{my-mnssm3}. The notation applied is $x: e, \mu, \tau, \gamma, q$ and $P$, $D$ stand for prompt and delayed, respectively.}
\begin{center}
\begin{tabular}{lll}
\br
 $Z$ decay & & $W^{\pm}$ decay\\
\mr
 $2 x^D 2 {\bar{x}}^D + \met\: (\text{via}~\tilde{\chi}_{i+3}\tilde{\chi}_{j+3})$ &
 & \multirow{2}{*}{$\ell^P/\tau^P + x^D {\bar{x}}^D + \met\: (\text{via}~\tilde\chi^\pm_i\tilde{\chi}_{j+3})$ }\\
 $2 x^P 2 {\bar{x}}^P \: (\text{via}~S^0_iP^0_j)$  & & \\
\br
\end{tabular}
\end{center}
\end{table}

These rare new decays are strongly constrained by the measurements of the $Z$ and $W$ total widths. Signatures  with $\tau$-leptons and/or $b$-jets and with displaced objects would be preferred when probing these decays in order to suppress the huge SM background. Their low production rate, e.g.~$BR\sim\mathcal{O}(10^{-5})$ for the $Z$, necessitates the high statistics that will become available when upcoming collides such as the \emph{GigaZ} and \emph{TeraZ} modes of the Linear Collider and TLEP, respectively, will be in operation~\cite{my-mnssm3}.

\section{Implications for dark matter}\label{sc:dm}

We address here the issue of (not necessarily cold) dark matter in SUSY models with $R$-parity violation. It has been shown that these seemingly incompatible concepts \emph{can} be reconciled in bRPV models with a gravitino~\cite{gravitino,martin,brpv-dm,brpv-split} or an axino~\cite{axino} LSP with a lifetime exceeding the age of the Universe. In both cases, RPV is induced by bilinear terms in the superpotential that can also explain current measurements on neutrino masses and mixings without invoking any GUT-scale physics. Decays of the next-to-lightest superparticle occur rapidly via RPV interactions, thus they do not disturb the Big-Bang nucleosynthesis, unlike the $R$-parity conserving case~\cite{leptog}. Decays of the NLSP into the gravitino and Standard Model particles do not contribute to the gravitino relic density in scenarios with broken $R$~parity. This is because decay processes involving a gravitino in the final state are suppressed compared to $R$-parity breaking decays unless the amount of $R$-parity violation is extremely small.

Interestingly, gravitino decays may produce monochromatic photons via $\t{G}\to\gamma\nu$, opening therefore the possibility to test this scenario with astrophysical searches, as we will see below. Requiring the model parameters to correctly account for observed neutrino oscillation parameters implies that expected rates for $\gamma$-ray lines produced by gravitino decays of mass above a few GeV would be confronted with the Fermi-LAT satellite observations, leading to an upper bound on the gravitino DM mass. For instance, the bRPV parameter region allowed by $\gamma$-ray line searches, dark matter relic abundance, and neutrino oscillation data, has been determined obtaining a limit on the gravitino mass $m_{\t{G}}\lesssim 1-10~\gev$ corresponding to a relatively low reheat temperature~\cite{brpv-dm}. The allowed region in the gravitino mass versus gravitino lifetime plane is shown in Fig.~\ref{fg:dm-brpv}. The yellow region is excluded by $\gamma$-ray line searches: Fermi and EGRET constraints are above and below 1~\gev, respectively.

\begin{figure}[htb]
\begin{minipage}{16pc}
\includegraphics[width=\textwidth]{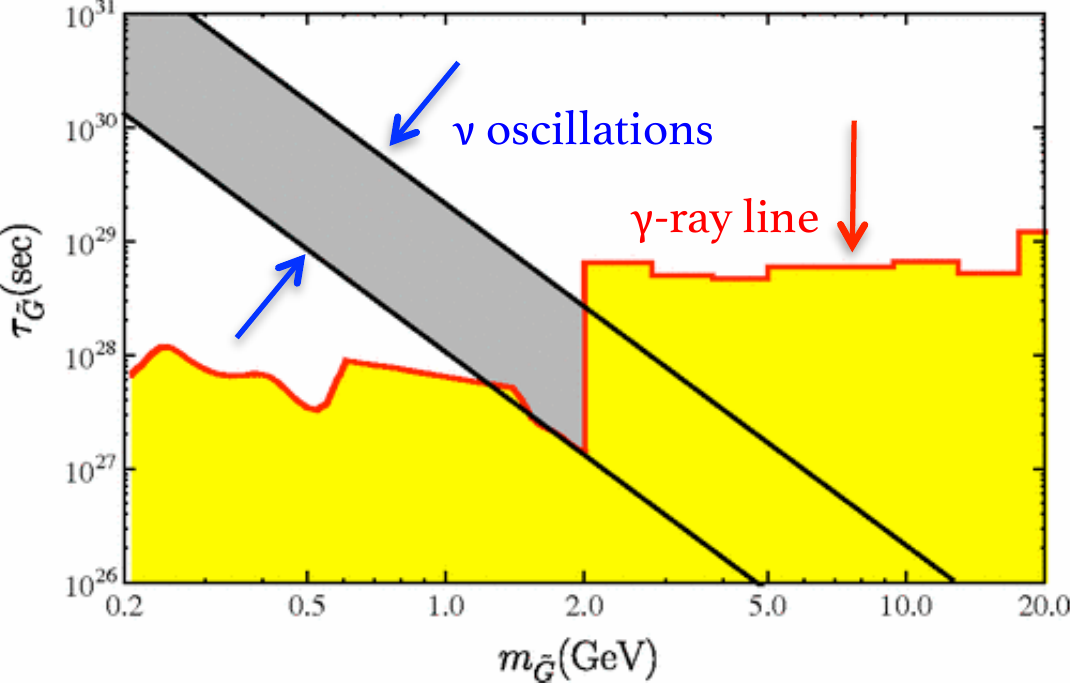}
\caption{\label{fg:dm-brpv}Allowed gravitino mass-versus-lifetime region (grey color) consistent with neutrino oscillation data and astrophysical bounds on $\gamma$-ray lines from dark matter decay for a bilinear RPV model. The lower and upper black lines correspond to $m_{1/2} = 240$ and $3000~\gev$, respectively. Adapted from Ref.~\cite{brpv-dm}. }
\end{minipage}\hspace{1.9pc}%
\begin{minipage}{20pc}
\includegraphics[width=\textwidth]{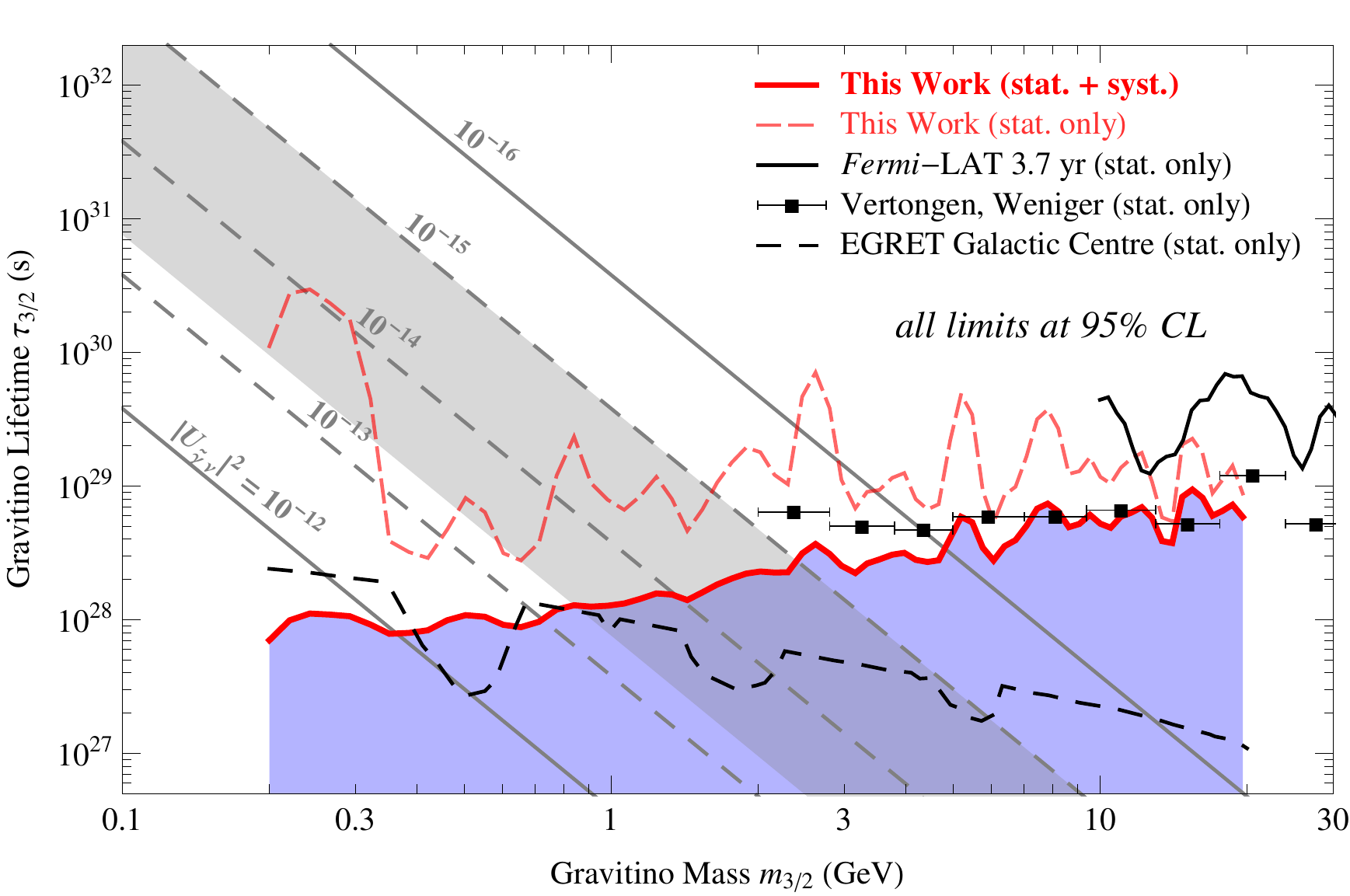}
\caption{\label{fg:fermi} Constraints on lifetime versus mass for decaying gravitino DM in the \mn. Diagonal grey lines (shaded band) denote allowed (favoured) photino-neutrino mixing parameters for the \mn, The blue shaded area is excluded by the limits set by 5-year Fermi-LAT data. The upper bounds set by several other $\gamma$-ray observations are also overlaid. From~\cite{mnssm-fermi}. }
\end{minipage} 
\end{figure}

Evidence on the four-year Fermi data that have found excess of a 130~\gev\ gamma-ray line from the Galactic Centre (GC)~\cite{fermi130} have been studied in the framework of $R$-parity breaking SUSY. A decaying axino DM scenario based on the SUSY KSVZ axion model with the bilinear $R$-parity violation explains the Fermi 130-\gev\ gamma-ray line excess from the GC while satisfying other cosmological constraints~\cite{brpv-axino}. On the other hand, gravitino dark matter with trilinear RPV ---in particular models with the $LLE$ RPV coupling--- can account for the gamma-ray line, since there is no overproduction of anti-proton flux, while being consistent with big-bang nucleosynthesis and thermal leptogenesis~\cite{trpv-gravitino}.

Measurements of the cosmic-ray antiproton flux by the PAMELA experiment have been used recently to constrain the (decaying) gravitino mass and lifetime in the channels $Z\nu$, $W\ell$ and $h\nu$~\cite{grefe}. Subsequently upper limits have been set on the size of the $R$-parity violating coupling in the range of $10^8$ to $8\times10^{13}$. These limits turn out to be more stringent than those coming from contributions to  neutrino masses or from the baryon asymmetry in the early Universe. Combining them with lower limits from big-bang nucleosynthesis constraints on the NLSP lifetime allows narrowing down the available parameter space for gravitino DM with bRPV.

Recent analyses of combined spectra of multiple galaxy clusters and the Andromeda galaxy from the XMM-Newton telescope, have revealed a tentative line with the central energy of 3.5~keV~\cite{xray35}. First and foremost, any long lived particle that produces enough number of photons should be a good candidate as a source of X-rays. Among the various theoretical scenarios attempting to explain it, RPV SUSY with an LSP decaying to photon and a neutrino is a possibility. In bilinear RPV it was found that a warm-dark-matter axino  with a mass of $m_{\tilde{\alpha}}\simeq7~{\rm keV}$ can have the proper lifetime and relic density to account for the observed X-ray emission line through its decay $\tilde{\alpha}\to\gamma\nu$~\cite{xray35-brpv}. Alternatively the line may be due to annihilating DM with $m_{\tilde{\alpha}}\simeq3.5~{\rm keV}$. Apart from the axino, other possible candidates include a gravitino, a bino, or a hidden sector photino as decaying DM leading to such a signal studied in the context of other-than-bilinear RPV SUSY scenarios~\cite{xray35-other}.

Such gravitino DM is also proposed in the context of \mn\ with profound prospects for detecting $\gamma$ rays from their decay~\cite{mnssm-gravitino1}. In studies on the prospects of the Fermi-LAT telescope to detect such monochromatic lines in five years of observations of the most massive nearby extragalactic objects, it was found that a gravitino with a mass range of $0.6-2~\gev$, and with a lifetime range of about $3\times10^{27} - 2\times10^{28}$~s should be detectable with a signal-to-noise ratio larger than three~\cite{mnssm-gravitino2}. After confronting the actual $\gamma$-ray flux data, limits on the model parameters have been set~\cite{mnssm-gravitino2,mnssm-fermi}. It was shown that gravitino masses larger than about 4~\gev\ are already excluded in the \mn\ by Fermi-LAT data of the galactic halo~\cite{mnssm-gravitino2}. The constraints on $\t{G}$ lifetime versus mass obtained by the 5-year Fermi-LAT and other $\gamma$-ray observations, as a consequence of the limits on line emission, are summarised in Fig.~\ref{fg:fermi}. If the photino-neutrino mixing parameter $|U_{\t{\gamma}\nu}|$ from the $7\times7$ neutralino mixing matrix is between $10^{-16}$ and $10^{-12}$, the correct neutrino masses are reproduced in \mn.  Values of the gravitino mass larger than 5~\gev,  are disfavoured, or lifetimes smaller than $\sim10^{28}~{\rm s}$, are excluded at 95\% CL as DM candidates.
 
$R$-parity breaking couplings can be sufficiently large to lead to interesting expectations for collider phenomenology. The neutralino NLSP, depending on the RPV model considered, may decay into~\cite{martin,ll-higgsinos}
\begin{eqnarray} \nonumber
& \X\to W^{\pm}\ell^{\mp},       & \qquad \X\to\t{G}\gamma, \\ 
& \X\to \nu\tau^{\pm}\ell^{\mp}, & \qquad \X\to\t{G}Z,      \\ \nonumber
& \X\to \nu\gamma,               &  \qquad \X\to\t{G}h.
\end{eqnarray}

Such decays may be probed at the LHC via inclusive channels characterised by leptons, many jets, large \met\ and/or photons or exclusive channels involving the reconstruction of a $Z$ or a $h$ from its decay products. The cases where the NLSP is long-lived yet with a decay length comparable to the dimensions of an LHC experiment are particularly interesting, as they give rise to displaced tracks/leptons and non-pointing photons. The possibility to measure the neutralino decay length provides an extra handle to constrain the underlying SUSY model. Nevertheless determining whether $R$~parity is conserved or broken may not be trivial since the DM particles themselves, whether absolutely stable or quasi-stable, cannot be detected directly in collider experiments. To this effect, an interplay between collider and astroparticle physics is necessary to pin down the dark matter properties and the related theoretical-model parameters.

\section{Summary and outlook}\label{sc:summary}

The hitherto null results of searches for supersymmetry in conventional channels calls for a more systematic and thorough consideration of non-standard SUSY theoretical scenarios and experimental techniques. To this effect, scenarios involving violations of $R$~parity and/or (meta)stable particles arise as interesting alternatives.

$R$-parity violating supersymmetry may reproduce correctly the measured neutrino physics parameters. Moreover its enriched mass spectrum and $R$-parity breaking decays can lead to novel signals at colliders, among which very few possibilities have been explored so far. In particular, searches for displaced objects at the LHC offer an attractive possibility as well as the prompt multilepton analyses. The former suffer from less background sources from SM processes when compared to prompt-object searches. More sophisticated searches are expected by ATLAS, CMS and LHCb in the near future during LHC Run~II, where energies of $13-14~\tev$ will be probed.

\ack

The author is grateful to the DISCRETE~2014 organising committee for the kind invitation to the Symposium. She thanks Pradipta Ghosh for providing material for the paper. She acknowledges support by the Spanish Ministry of Economy and Competitiveness (MINECO) under the project FPA2012-39055-C02-01, by the Generalitat Valenciana through the project PROMETEO~II/2013-017 and by the Spanish National Research Council (CSIC) under the JAE-Doc program co-funded by the European Social Fund (ESF). 


\end{document}